\documentclass[12pt]{article}
\usepackage{graphicx}
\usepackage{amssymb}
\usepackage{amsmath}

\newcommand{\ha}{H_\alpha}
\newcommand{\gab}{g_{\alpha\beta}}
\newcommand{\gba}{g_{\beta\alpha}}
\newcommand{\la}{\Lambda_\alpha}

\newcommand{\ra}{\rho_\alpha}

\newcommand{\aaa}{A_\alpha}
\newcommand{\vni}{\vec{n}_i}
\newcommand {\Ang}{\mbox{\AA}}

\begin{document}

\baselineskip17pt

\title{How Events Come Into Being: EEQT, Particle
Tracks, Quantum Chaos, and Tunneling Time}
\author{Ph.~Blanchard, A.~Jadczyk\footnote{On leave from: Institute of Theoretical
Physics, University of Wroclaw. Now with: Constellation Technology
Corporation, Largo, FL 33777-1406, e-mail: ajad@ift.uni.wroc.pl},
A.~Ruschhaupt\\ Faculty of Physics and BiBoS\\ University of
Bielefeld\\ Universit\"atstr. 25\\ D-33615 Bielefeld}

\date{}

\maketitle
\begin{abstract}
In sections 1 and 2 we review Event Enhanced Quantum Theory
(EEQT). In section 3 we discuss applications of EEQT to tunneling
time, and compare its quantitative predictions with other
approaches, in particular with B\"uttiker-Larmor and Bohm
trajectory approach. In section 4 we discuss quantum chaos and
quantum fractals resulting from simultaneous continuous monitoring
of several non-commuting observables. In particular we show
self-similar, non-linear, iterated function system-type, patterns
arising from quantum jumps and from the associated Markov
operator. Concluding remarks pointing to possible future
development of EEQT are given in section 5.
\end{abstract}

\section{Introduction}
 Event-Enhanced Quantum Theory (EEQT) was developed in response to
John Bell's concerns about the status of the measurement problem
in quantum theory \cite{bell89,bell90}. The main thrust of quantum
measurement theory is to explain the mechanism by which potential
properties of quantum systems become actual. At the present time,
this is no longer an abstract or philosophical problem since it is
now possible to carry out prolonged observations of individual
quantum systems. These experiments provide us with time series
data, and a complete theory must be able to explain the mechanism
by which these time series are being generated; moreover it must
be in position to ``simulate" the process of events generation.

John Bell sought a solution to the measurement problem in hidden
variable theories of Bohm and Vigier, his own idea of beables, and
also in the spontaneous localization idea of Ghirardi, Rimini and
Weber \cite{grw}. More recently we have proposed a formalism going
in a similar direction, but avoiding the introduction of hidden
variables beyond the wave function itself
\cite{blaja93a,blaja95a,blaja95b}.

EEQT offers a mathematically consistent way of coupling between a
quantum and a classical system. The classical system $C$ is
described by an Abelian algebra ${\cal A}_c$. In this respect,
EEQT is indeed an enhancement because it modifies the quantum
dynamics by adding a new term to the Liouville equation. This
allows unification of the continuous evolution of quantum states
with quantum jumps that accompany real world events. When the
coupling $Q-C$ is weak, events are sparse and EEQT reduces to the
standard quantum theory.

In this EEQT framework, a measurement process is a coupling $Q-C$,
where transfer of information about the quantum state to the
classical recording device is mathematically modeled by a
semigroup of completely positive and trace-preserving maps of the
total system $Q\times C$. Let us emphasize that such a transfer of
information cannot, indeed, be implemented by a Hamiltonian or
more generally by any automorphic evolution \cite{lands91,jad94a}.

To illustrate this last point, let us consider the total system
$Q\times C$ described by an algebra of operators ${\cal A}$ with
center $z = {\cal A}_c$. The center describes the classical
degrees of freedom. Let $\omega$ be a state of ${\cal A}$ and let
$\omega_c$ denote its restriction to ${\cal A}_c$. Let $\alpha_t$
be an automorphic time evolution of ${\cal A}$ and $\omega_t =
\alpha_t (\omega)$ the dual evolution of states given by
$\alpha^t_t(\omega) (A) = \omega (\alpha_t (A))$. Each
automorphism of $\cal A$ leaves its center invariant, which
implies that $\alpha_t (\omega)_c = \alpha^t |_z (\omega_c)$. In
other words $\omega^t_c$ depends only on $\omega^0_c$ and the
state at time $t$ of $C$ depends only on the part of $C$ and not
on its extension to the total system $Q\times C$. The result is
that to have information transfer from the total system to its
classical subsystem we must use non automorphic time evolutions.

The formal development of EEQT was inspired by the works of Jauch
\cite{jauch64}, Hepp \cite{hepp72}, Piron
\cite{piron76,piron83,piron81}, Gisin \cite{gispir81,gisin84},
Araki \cite{araki85} and Primas \cite{primas81,primas90}. In
\cite{davis84,davis92} M.H.A.~Davis described a special class of
piecewise deterministic Markov processes that reproduced the
master equation postulated in \cite{blaja93a}. This opened a new
chapter of EEQT and allowed for description of individual quantum
systems. In \cite{jakol95} it was proven that the class of
couplings considered in EEQT leads to a unique piecewise
deterministic Markov process taking values on the pure state space
of the total system $Q\times C$. This process consists of random
jumps accompanied by changes of the classical state, interspersed
by random periods of Schr\"odinger like deterministic evolution.
The process is nonlinear in the quantum pure state $\psi$ and
after averaging we recover the original linear master equation for
statistical states of the total system $Q\times C$.

The crucial concept of EEQT is that of a classical, discrete and
irreversible event. This is taken into account by including, from
the beginning, classical degrees of freedom. Once the existence of
the classical part is accepted then ``events" can be defined as
changes of pure state of this classical part $C$. In EEQT events
do happen and they do it in finite time. Rudolf Haag \cite{haag96}
takes a similar position and calls it an ``evolutionary picture".
According to this view the future does not yet exist and is being
continuously created, this creation being marked by events.

In EEQT we have a flow of information from $Q$ to $C$ and moreover
a way to calculate numbers in real experiments and to model the
feedback from $C$ to $Q$. The coupling $Q\times C$ does not mean
we are taking a step backward into classical mechanics. We are
only claiming that not all is quantum and that there are elements
of Nature that are not, and cannot be, described by a quantum wave
function. This assumption is confirmed everyday by experiments
which clearly show that we are living in a world of facts and not
in a world of potentialities. For this aspect which is not
reducible to quantum degrees of freedom we use the term "classical
variables". This does not imply that we impose any restriction on
their nature.

At this point we would like to emphasize a fundamental difference
between the classical variables of EEQT and the additional
parameters introduced in hidden variable theories. Hidden variable
theories consider microscopic variables that are hidden from our
observation. EEQT deals with classical variables that can be
directly observed. They are a direct counterpart of Physics on the
other side of the Heisenberg-von Neumann cut. Another important
point is that in hidden variable theories there is no back action
of these variables on the wave function. In EEQT we have a
feedback of $C$ on $Q$. EEQT can be also considered as a final
result of a decoherence mechanism as described in
\cite{omnes99,zurek98}. In section 2 the mathematical formalism of
EEQT is presented. In sections 3 and 4 some applications are
described. Concluding remarks are given in section 5.

\section{EEQT -- The mathematical formalism}

We will only describe the case of a discrete classical system $C$.
It has been shown in \cite{blaja93c}, while applying EEQT to
SQUID, how to extend the formalism in cases where the classical
system $C$ is continuous.   There are two levels in EEQT - the
ensemble level and the individual level. This is in a total
contradistinction to the standard quantum theory which deals only
with ensembles and even claims, rather often, that individual
description is impossible! Let us begin with the ensemble level.\\
First of all, in EEQT, at that level, we use all the standard
mathematical formalism of quantum theory, but we extend it adding
an extra, possibly multidimensional, parameter $\alpha .$ Thus all
quantum operators $A$ get an extra index $A_\alpha$, quantum
Hilbert space ${\cal H}$ is replaced by a family ${\cal H}_\alpha
,$ quantum state vectors $\psi$ are replaced by families
$\psi_\alpha ,$ quantum Hamiltonian $H$ is replaced by a family
$H_\alpha$ etc.\\ The parameter $\alpha$ is used to distinguish
between macroscopically different and non-superposable states of
the universe. In the simplest possible model we are interested
only in describing a "yes-no" experiment and we disregard any
other parameter - in such a case $\alpha$ will have only two
values $0$ and $1$. Thus, in this case, we will need two Hilbert
spaces. This will be the case when we will deal with sharp
particle detectors. In a more realistic situation $\alpha$ will
take values in a multi-dimensional, perhaps even
infinite-dimensional manifold. But even that may prove to be
insufficient.

When, for instance, EEQT is used as an engine powering
Everett-Wheeler many-world branching-tree, in  such a case,
$\alpha$ will also have to have the corresponding dynamical
branching-tree structure, where the space in which the parameter
$\alpha$ takes values, grows and becomes more and more complex
together with the growing complexity of the branching structure.

An ``event" is, in our mathematical model, represented by a change
of $\alpha$, $\alpha$ representing a pure state of the classical
subsystem $C$. This change is discontinuous, is a branching.
Depending on the situation this branching is accompanied by a more
or less radical change of physical parameters. Sometimes, such as
in the case of a phase transition in Bose-Einstein condensate, we
will need to change the nature of the underlying Hilbert space
representation. In other cases, such as the case of a particle
detector, the Hilbert spaces ${\cal H}_0$ and ${\cal H}_1$ will be
indistinguishable copies of one standard quantum Hilbert space
${\cal H}.$

Another important point is this: time evolution of an individual
quantum system is described by piecewise continuous function
$t\mapsto\alpha(t)$, $\psi(t)\in{\cal H}_\alpha(t)$, a trajectory
of a piecewise deterministic Markov process (in short: PDP), where
periods of continuous evolution are interspersed by discontinuous
catastrophic jumps.

As already pointed out above, in EEQT any non-trivial coupling of
classical to quantum degrees of freedom involves back-action of
classical on quantum. This back-action shows up in a dual way: in
changes to the continuous evolution (as in "interaction free
measurements") and also in discontinuous jumps and branchings. It
is impossible to understand the essence of this back-action
without having even a rough idea about PDPs.

Originally EEQT was described in terms of a master equation for a
coupled, quantum+classical, system; thus it was only applicable to
ensembles; the question of how to describe individual systems was
open. Then, after searching through the mathematical literature,
we found that, in his monographs \cite{davis84,davis92} dealing
with stochastic control and optimization, M. H. A. Davis described
a special class of piecewise deterministic processes that fitted
perfectly the needs of quantum measurement theory, and that
reproduced the master equation postulated originally by us in Ref.
\cite{blaja93a}. The special class of couplings between a
classical and quantum system leads to a unique piecewise
deterministic process with values on $E$-the pure state space of
the total system. This process consists of random jumps,
accompanied by changes of a classical state, interspersed by
random periods of Schr\"odinger-type (but non-unitary)
deterministic evolution. The process, although mildly nonlinear in
quantum wave function $\psi$, after averaging, recovers the
original linear master equation for statistical states.\\

We would like to stress that, in EEQT, the dynamics of the coupled
total system which is being modeled is described not only by a
Hamiltonian ${\cal H}$, or better: not only by an $\alpha$--
parametrized family of Hamiltonians $H_\alpha$, but also by a
doubly parametrized family of operators $\{g_{\beta\alpha}\}$,
where $g_{\beta\alpha}$ is a linear operator from ${\cal
H}_\alpha$ to ${\cal H}_\beta$. While Hamiltonians must be
essentially self-adjoint, $g_{\beta\alpha}$ need not be such --
although in many cases, when information transfer and control is
our concern (as in quantum computers), one wants them to be even
positive operators (otherwise unnecessary entropy is created).

It should be noted that the time evolution of statistical
ensembles is, due to the presence of $\{g_{\beta\alpha}\}$'s, {\sl
non-unitary} or, using algebraic language, {\sl non-automorphic}.
The system, as a whole, is open. This is necessary, as we like to
emphasize: information (in this case: information gained by the
classical part) must be paid for with dissipation!  There is no
free lunch!

A general form of the linear master equation describing
statistical evolution of the coupled system is given by
\begin{equation}
{\dot A}_\alpha=i[\ha,\aaa]+\sum_\beta \gba^\star A_\beta \gba -
{1\over2}\{\la,\aaa\},
\end{equation}
\begin{equation}
{\dot \rho}_\alpha=-i[\ha,\ra]+\sum_\beta \gab \rho_\beta
\gab^\star - {1\over2}\{\la,\ra\},
\end{equation}
where
\begin{equation}
\la=\sum_\beta \gba^\star \gba.
\end{equation}
The operators $\gab$ can be allowed to depend explicitly on time.
While the term with the Hamiltonian describes "dyna-mics", that is
exchange of forces, of the system, the term with $\gab$ describes
its "bina-mics" - that is exchange of "bits of information"
between the quantum and the classical subsystem.\\

As has been proven in \cite{jakol95} the above Liouville equation,
provided the diagonal terms $g_{\alpha\alpha}$ vanish, can be
considered as an average of a unique Markov process governing the
behavior of an individual system. The real--time behavior of such
an individual system is given by a PDP process realized by the
following non--unitary, non--linear and non--local, EEQT
algorithm:

\medskip
\noindent {\bf PDP Algorithm} {\it Suppose that at time $t_0$ the
system is described by a quantum state vector $\psi_0$ and a
classical state $\alpha$. Then choose a uniform random number
$p\in [0,1]$, and proceed with the continuous time evolution by
solving the modified Schr\"odinger equation $$ {\dot
\psi_t}=(-iH_\alpha dt-{1\over2}\la )\psi_t $$ with the initial
wave function $\psi_0$ until $t=t_1$, where $t_1$ is determined by
$$\int_{t_0}^{t_1} (\psi_t,\la \psi_t ) dt = p.$$ Then jump. When
jumping, change $\alpha\rightarrow \beta$ with probability
$$p_{\alpha\rightarrow\beta}=
\Vert\gba\psi_{t_1}\Vert^2/(\psi_{t_1}\la,\psi_{t_1}),$$ and
change $$\psi_{t_1}\rightarrow\psi_1=\gba\psi_{t_1}/
\Vert\gba\psi_{t_1}\Vert.$$

Repeat the steps replacing $t_0,\psi_0,\alpha$ with
$t_1,\psi_1,\beta$.}

\medskip
\noindent The algorithm is non--linear, because it involves
repeated normalizations. It is non-unitary because of the extra
term $-{1\over2}\la $ in the exponent of the continuous evolution.
It is non--local because it needs repeated computing of the norms
- they involve instantaneous space--integrations. It is to be
noted that PDP processes are more general than the popular
diffusion processes. In fact, every diffusion process can be
obtained as a limit of a family of PDP processes.

\section{Cloud chamber model, GRW Spontaneous Localization Theory
and Born's Interpretation}

In this example, we wish to account for the tracks that quantum
particles leave in cloud chambers. Physically a cloud chamber is a
highly complex system. To describe the response of the chamber to
a quantum particle it is sufficient to assume that we have to deal
with a collection of two state systems able to change their state
when a particle passes near a sensitive center. Let us sketch the
model proposed in \cite{jad94b,jad94c}.

Let us consider the space $E = {\bf R}^3$ as filled with a
continuous medium (photographic emulsion, super-saturated vapor,
etc.) which can be at each point $a\in E$ in one of two states:
``on" represented by ${1 \choose 0}$ and ``off" represented by
${0\choose 1}$. The set of all possible states of the system is
then $2^E$. But we are only interested with a continuum of states
- namely the ``vacuum" (i.e. when all points of the medium are in
``off" state)- and states which differ from the vacuum only in a
finite number of points. We define "event" to be a change of state
of a finite number of points. Thus the space of classical events
can be identified with the space of finite subsets of $E$ from
which it follows that the total system $\Sigma_{tot} = \Sigma_q
\otimes \Sigma_c$ is described by families $\{
\rho_\Gamma\}\subset E$, $\Gamma$ finite subset of $E$. For each
$a\in E$ let $g_a$ be a Hermitian bounded operator which
represents heuristically the sensitivity of the two-state detector
located at $a$. We can think of $g_a$ as a gaussian function $g_A
(x)$ centered at $x = a$ (other phenomenological shapes are also
possible). We denote
\begin{equation}
\int_E g^2_a (x) da = \Lambda (x)\ . \end{equation} The quantum
mechanical Hilbert space is then ${\cal H}_q = L^2 ({\bf R}^3,
dx)$. Each state $\rho$ of the total system can be, formally,
written as $\rho = \Sigma_{\Gamma\in {\cal S}} \rho_\Gamma \otimes
\epsilon_\Gamma$, where, for $\Gamma\in{\cal S}$,
\begin{equation}
\epsilon_\Gamma = \prod \otimes_{a\in E} \left( \begin{array}{cc}
\chi_\Gamma(a) & 0\\ 0 & 1 - \chi_\Gamma (a))\end{array} \right)\,
\end{equation}
and where $\chi_\Gamma$ stands for the characteristic function
of $\Gamma$. The Lindblad coupling is now chosen in the following
way
\begin{equation}
L_{int} (\rho) \equiv \int_{{\bf R}^3} da \left[ V_a \rho V_a -
\frac{1}{2} \{ V^2_a, \rho \} \right] \end{equation} where $V_a =
g_a \otimes \tau_a, \tau_a$ denoting the flip at the point $a \in
{\bf R}^3$. Let us introduce the following notation: $a(\Gamma)$
represents the state $\Gamma$ with the counter at position $a$
flipped, i.e. $a(\Gamma) = (\Gamma\backslash\{ a\} \cup \{ \{ a
\}\backslash \}$. The Liouville equation is given by $\dot{\rho} =
- i \left[ H,\rho\right] + L_{int} (\rho)$. But using the
following identity in eq. (6)
\begin{equation}
V_a\rho V_a = \sum_\Gamma g_a \rho_\Gamma g_a \otimes
\epsilon_{a(\Gamma)} = \sum_\Gamma da g_a \rho_{_{a(\Gamma)}} g_a
- \frac{1}{2} \{ \Lambda, \rho_{\Gamma} \otimes \epsilon_a\}_{+}
\end{equation} we can write
\begin{equation}
\dot{\rho}_\Gamma = - i \left[ H, \rho_\Gamma\right] + \int_{{\bf
R}^3} da g_a \rho_{_{a(\Gamma)}} g_a - \frac{1}{2} \{ \Lambda,
\rho_\Gamma \} \ . \end{equation} Summing up over $\Gamma$ we get
for the effective quantum state $\hat{\rho} = \Sigma_\Gamma
\rho_\Gamma$
\begin{equation}
\dot{\hat{\rho}} = - i\left[ H, \hat{\rho}\right] - \int_{{\bf
R}^3} dag_a \hat{\rho}g_a - \frac{1}{2} \{ \Lambda,\hat{\rho}\}\ .
\label{eq:grwa} \end{equation}

Let us emphasize that the time derivative of $\hat{\rho}$ depends
only on $\hat{\rho}$. Moreover the effective Liouville equation is
exactly of the type discussed in connection with the spontaneous
localization model of Ghirardi, Rimini and Weber \cite{grw}, the
difference being that GRW considered only the constant rate case,
and were simply not interested in the classical traces of
particles. Indeed if, following GRW, we take for $g_a$ the
Gaussian functions:
\begin{equation}
g_a(x)=\sqrt{\frac{\lambda}{2}}\left(\frac{\alpha}{\pi}\right)^{3\over
4}
 \exp\left({-\frac{\alpha (x-a)^2}{2}}\right). \end{equation} then
$\Lambda(x)\equiv\frac{\lambda}{2}$ and Eq.(\ref{eq:grwa}) becomes
\begin{equation}
{\dot {\hat\rho}} = -i[H,{\hat\rho}]+\int da\, g_a{\hat\rho} g_a
-\lambda\, {\hat\rho },\end{equation} exactly as in GRW [3]. Thus
we have\\

\noindent {\bf Theorem GRW:} {\sl Ghirardi--Rimini--Weber
spontaneous localization model is an effective quantum evolution
part of a particular case of EEQT type coupling of a quantum
particle to a homogeneous two-state classical detector medium.}\\

We can also construct the associated PD Markov process. We get for
time evolution observables the same equation as in (8) except for
the sign in front of the Hamiltonian. By taking expectation values
we obtain a Davis generator corresponding to rate function
$\lambda (\psi) = (\psi, \Lambda\psi)$, and probability kernel $Q$
with non-zero elements of $Q$ given by
\begin{equation}
Q(\psi, \Gamma; d\psi^\prime, a(\Gamma)) = \frac{\parallel
g_a\psi\parallel^2}{\lambda(\psi)} \delta\left(\psi^\prime -
\frac{g_a \psi}{\parallel g_a\psi\parallel}\right)  d\psi^\prime\
. \end{equation} Time evolution between jumps is given by:
\begin{equation}
\psi_t = \frac{\exp (-iHt - \frac{\Lambda t}{2})\psi_0}{\parallel
\exp (-iHt - \frac{\Lambda t}{2}\psi_0)\parallel}\ .
\end{equation}

The PD process can be described as follows: $\psi \in L^2 ({\bf
R}^3,dx)$ develops according to the above formula until at time
$t_1$ jump occurs. The jump consists of a pair: (classical event,
quantum jump). The classical medium jumps at $a$ with probability
density
\begin{equation}
p(a; \psi_{t_1}) = \parallel g_a \psi_{t_1}
\parallel^2/\lambda(\psi_{t_1})\ , \end{equation} (flip of the detector)
while the quantum part of the jump is jump of the Hilbert space
vector $\psi_{t_1}$ to $g_a\psi_{t_1}/\parallel
g_a\psi_{t_1}\parallel$ and the process starts again. The random
jump time $t_1$ is governed by the inhomogeneous Poisson process
with rate function $\lambda (\psi_t)$. If the medium is
homogeneous, then $\lambda(\psi) =$ const $= \lambda$, and we
obtain for quantum jumps the GRW spontaneous localization model.
More complete discussion can be found in refs. [24,25].

\bigskip
\medskip
\noindent {\bf Derivation of Born's interpretation} \ \ Let us
consider now the idealized case of a homogeneous medium of
particle detectors that are coupled to the particle only for a
short time interval $(t , t + \Delta t)$, $\Delta t \to 0$ with
intensity $\lambda$, so that $\lambda\Delta t \to \infty$. Let us
also assume that the detectors are strictly point-like that is,
that $g^2_a (x) \to \lambda\delta(x-a)$. In this case the formula
(12), giving the probability density of firing the detector at
$a$, becomes $p(a; \psi) = \parallel g_ \psi \parallel^2/\lambda =
|\psi(a)|^2$ and we recover the Born interpretation of the wave
function. The argument above goes as well for the case of a
particle with spin.

\section{Tunneling time}
In this section we will discuss, in the EEQT framework, the
following questions: How long is the mean reflection time, that
is, the mean time which an electron spends in the barrier if
reflected? How long is the mean traversal time; the mean time an
electron spends in the barrier if transmitted?

Therefore an operational definition of traversal and reflection
times is used similar to the approach of Palao, Muga, Brouard and
Jadczyk \cite{palao.1997}, but we will examine both traversal and
reflection times.

\begin{figure}[p]
  \begin{center}
    \leavevmode
    \includegraphics [width=.9\linewidth, height=6cm]{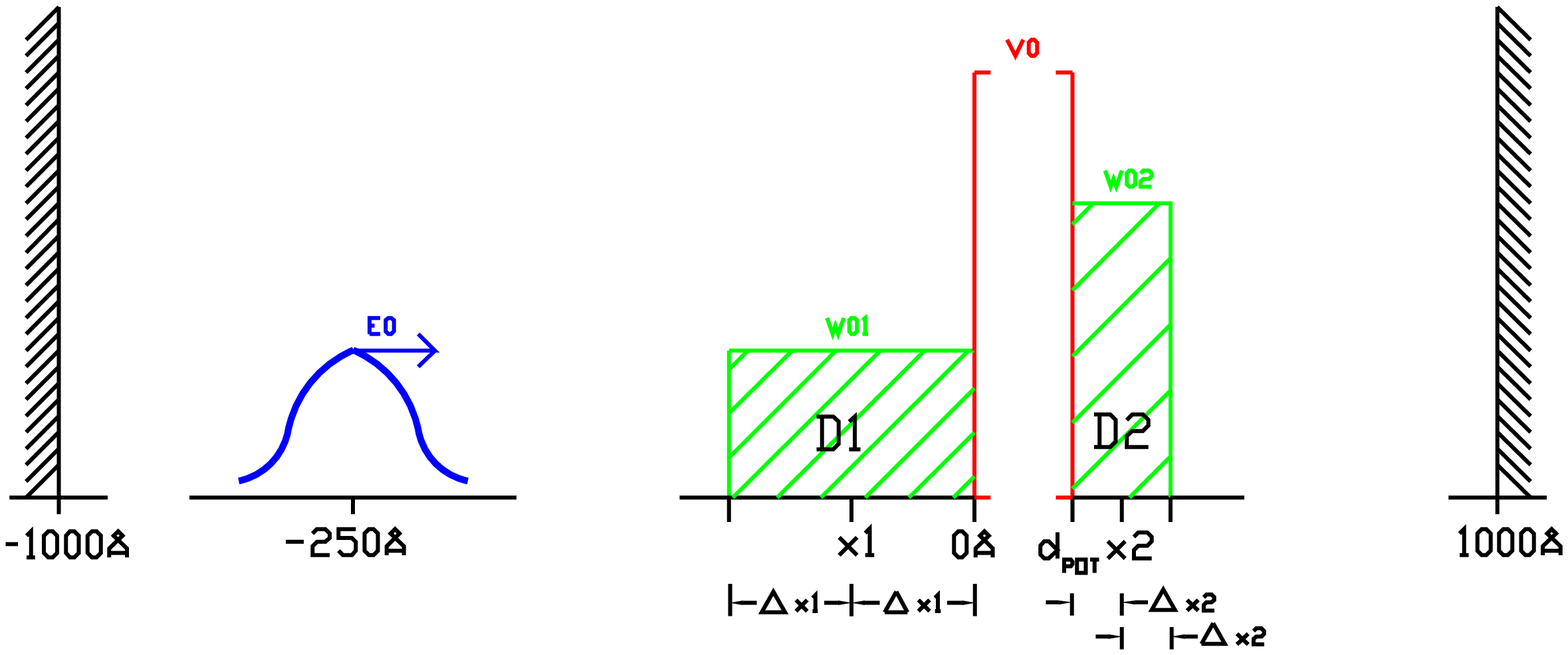}
  \end{center}
  \caption{Simulation situation} \label{fig_1}
\end{figure}

Let us consider the situation in one dimension (Fig.~\ref{fig_1}),
the potential is given by
\begin{eqnarray}
  V (x) = \left\{\begin{array}{r@{\quad:\quad}l} V_0 & 0 \le x \le
  d_{POT} \\ 0 & \mbox{otherwise}
\end{array} \right.
\end{eqnarray}
$d_{POT}$ being the width of the barrier.

A detector $D_1$ is put in front of the barrier which can detect
the particle without destroying it. A second detector $D_2$ should
be put behind the barrier.

At the beginning only the detector $D_1$ is active. When it
detects the particle at a time $t_0$, it turns on the detector
$D_2$ while keeping itself turned on. So the particle must be
detected first by $D_1$ (the possibility, that $D_2$ detects the
particle before $D_1$ is therefore avoided). Thus the particle can
be detected a second time by the detector $D_1$ or the detector
$D_2$. If the detector $D_1$ detects the particle a second time at
time $t_1$, the time difference $t_1 - t_0$ is defined as the
reflection time $t_{REF}$. If the particle is detected by the
detector $D_2$ at a time $t_2$, then the time difference $t_2 -
t_0$ is by definition the traversal time $t_{TRA}$.

Another possibility is, that the particle is never detected or is
detected only once. Therefore the experiment or simulation should
be stopped after a finite time $t_{CUT}$. The above definitions of
the traversal and reflection times are of course positive and
real.

A single run of the above experiment can be simulated by using the
PDP-algorithm of the EEQT (more details can be found in
\cite{ruschhaupt.1998}). Making a lot of runs, the mean reflection
time $\tau_{R,SIM}$ is given by averaging all runs which result in
a reflection time $t_{REF}$ and the mean traversal time
$\tau_{T,SIM}$ is given by averaging all times of runs, which
yield a traversal time $t_{TRA}$.

There exist many other approaches for calculating mean traversal
and reflection times of the interval $[x_1,x_2]$ between the
centers of the two detectors (for a review see for example
\cite{hauge.1989,landauer.1994,nimtz.1997}).

One of these are the phase time introduced by Hartman
\cite{hartman.1962}, which is similar to following the peak of the
wave packet or the ''semi-classical'' time, which is derived out
of the classical expressions.

Another possibility is to install an infinitesimally small
magnetic field in the range $[x_1,x_2]$ and look at the rotation
angle of the electron spin, this is the idea of the
B\"uttiker-Larmor traversal time derived by B\"uttiker
\cite{buettiker.1983}.

In the Bohm trajectory approach, one can talk about trajectories
of particles and therefore there exists a clear definition of
traversal and reflection times (for example see
\cite{leavens.1990a,leavens.1990b,leavens.1995a,leavens.1995b,oriols.1996}).

The mean reflection time $\tau_{R,SIM}$ computed by the
PDP-algorithm of the EEQT is compared which the results of the
above approaches (the exact formulas of the results in the other
approaches can be found in \cite{ruschhaupt.1998}).

\begin{figure}[p]
  \begin{center}
    \leavevmode
    \includegraphics [width=0.9\linewidth]{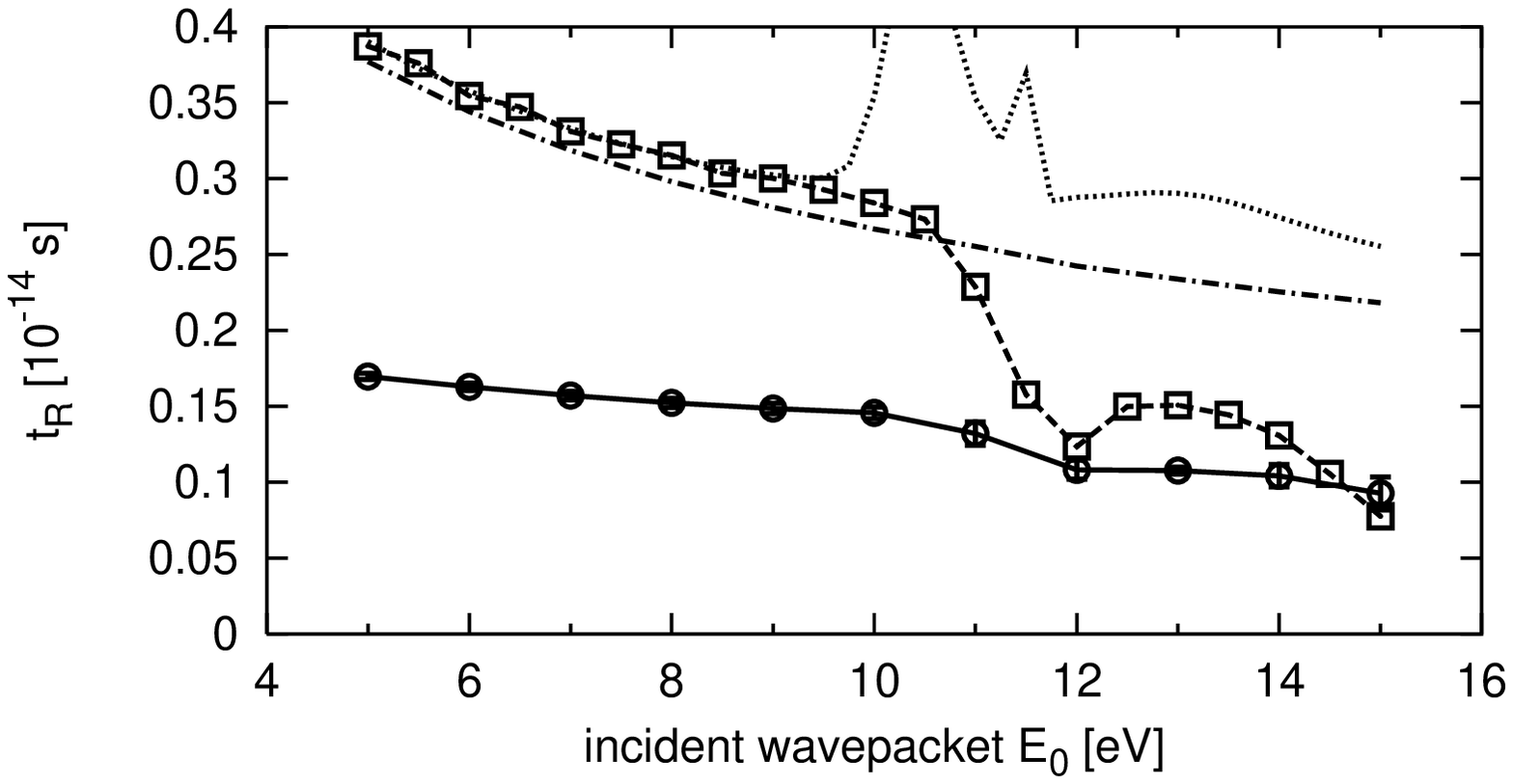}
  \end{center}
    \caption{Mean reflection time versus particle energy $E_0$}
    {simulation parameters: initial wave packet: $x_0 = -250
    \Ang$,
      $\eta=25\Ang$, barrier: $d_{POT} = 5 \Ang$, $V_0 = 10 eV$,
      detector $D_1$: $x_1 = - 25\Ang$, $\Delta x_1=25\Ang$,
      $W_{01} = 0.16 eV$, detector $D_2$: $x_2 = 10\Ang$, $\Delta
      x_2 = 5\Ang$, $W_{02}=2.56 eV$;

      mean reflection time $\tau_{SIM,R}$ (solid line with circles
      and errorbars);

      phase time approach : wave packet (dotted line);

      ``semi-classical'' reflection time : wave packet
(dashed-dotted line);

      Bohm trajectory approach (boxes with dashed line)}
    \label{fig_2}
\end{figure}

One result is, that the mean reflection time $\tau_{R,SIM}$ is
mostly smaller than those of the other
approaches(Fig.~\ref{fig_2}). The reason is that the first
detector cannot distinguish whether the particle is traveling
toward the barrier, or is returning from the barrier, when it is
detected a second time. So reflection times of particles, which do
not reach the barrier, are also measured.

Another question is: how the mean traversal time $\tau_{T,SIM}$
depends on the barrier length. The phase time results for plane
waves are independent of the barrier length. This fact is called
the ''Hartman-effect''. This effect was also seen in experiments
with photons (for example the experiments done by Enders and Nimtz
\cite{enders.1992,enders.1993c,enders.1993b}, done by Steinberg,
Kwiat and Chiao \cite{steinberg.1993} and done by Spielmann,
Szip\"ocs, Stingl and Krausz \cite{spielmann.1994}).

Here electrons are used and an additional detector is put before
the barrier in contrast to the photon-experiments. The question
then is, whether there is still a ''Hartman-effect'' or if there
is no such effect due to the fact of the additional detector.

\begin{figure}[p]
  \begin{center}
    \leavevmode
    \includegraphics [width=0.9\linewidth]{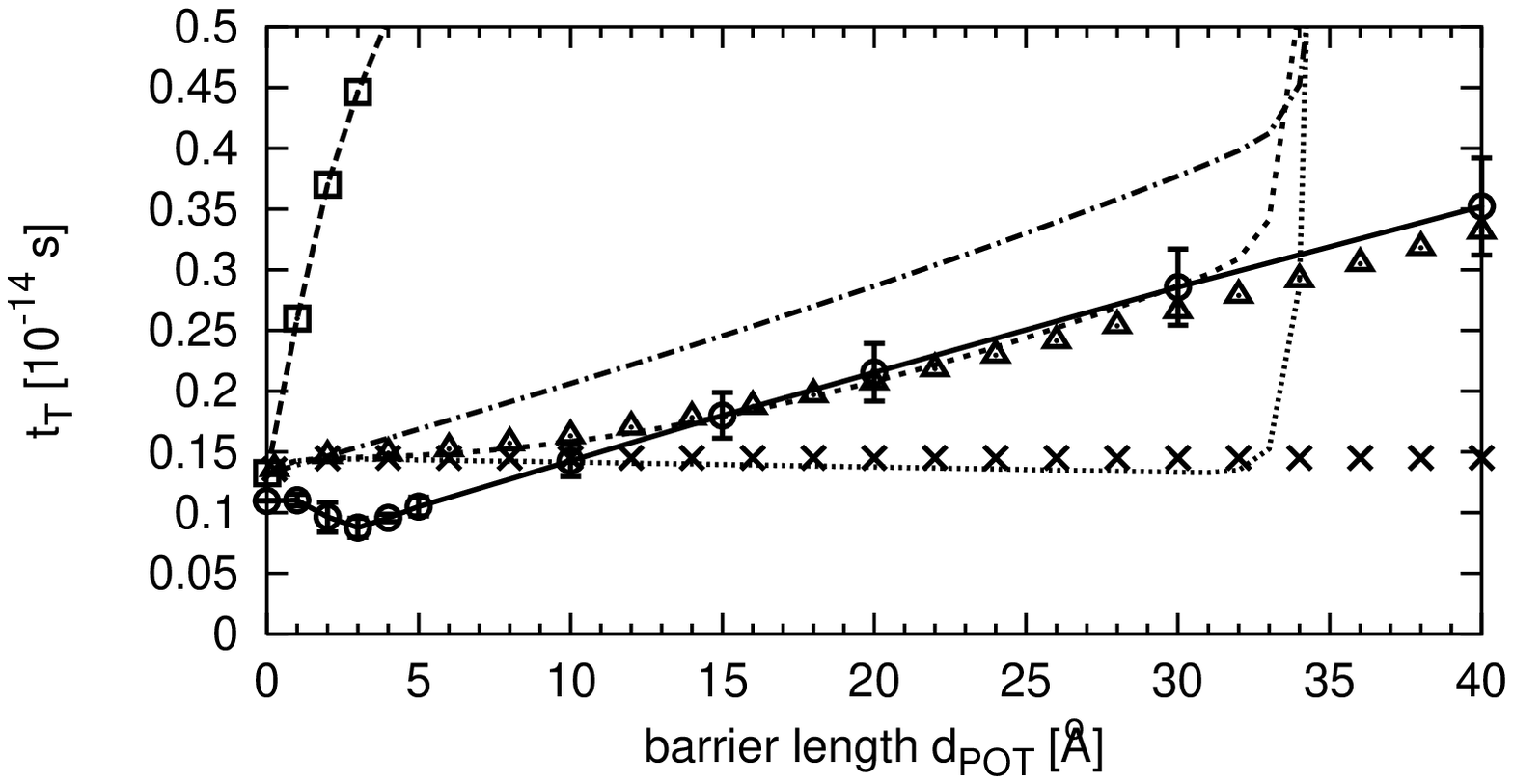}
  \end{center}

  \begin{center}
    \leavevmode
    \includegraphics [width=0.9\linewidth]{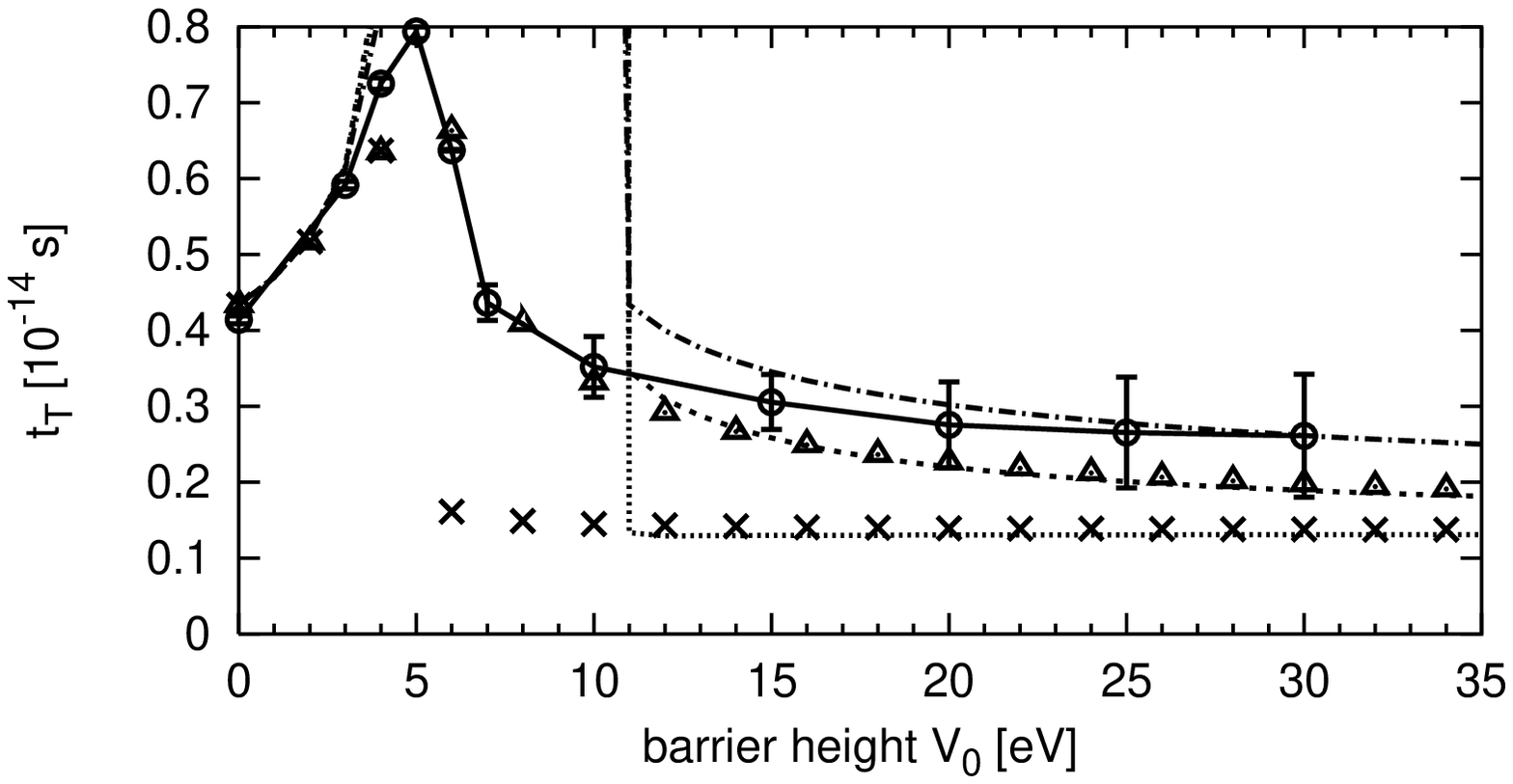}
  \end{center}

    \caption{Mean traversal time} {simulation situation: initial
    wave packet: $E_0 = 5 eV$, $x_0 = -250 \Ang$,
      $\eta=12.5 \Ang$, detector $D_1$: $x_1 = - 12.5 \Ang$,
      $\Delta x_1=12.5\Ang$, $W_{01} = 0.16 eV$, detector $D_2$:
      $x_2 = d_{POT} + 5\Ang$, $\Delta x_2 = 5 \Ang$, $W_{02}=2.56
      eV$,

      mean traversal time $\tau_{SIM,T}$ (solid line with circles
      and errorbars);

      phase time approach: plane wave (crosses), wave packet
(dotted line);

      ``semi-classical'' traversal time: wave packet
(dashed-dotted line);

      B\"uttiker Larmor Time: plane wave (triangles), wave packet
      (small-dashed line);

      Bohm Trajectory approach (boxes with dashed line)

      (a) versus barrier width $d_{POT}$, $V_0 = 10 eV$

      (b) versus barrier height $V_0$, $d_{POT} = 40 \Ang$}
    \label{fig_3a_3b}
\end{figure}

In our simulations, the energy of the particle, the height of the
barrier, and the detector parameters are fixed and the barrier
length $d_{POT}$ is varied (Fig.~\ref{fig_3a_3b}(a)).

The simulated times $\tau_{T,SIM}$ grow with increasing barrier
width $d_{POT} \ge 3\Ang$.  With these detectors the simulation
shows no wider range of barrier width with constant traversal
times, and therefore no ''Hartman-effect''.

The simulation results and the Larmor clock results for plane
waves show qualitatively the same characteristic: nearly the same
linear growth with increasing barrier width.

Another result is: how the mean traversal time depends on the
barrier height for very wide barriers (Fig.~\ref{fig_3a_3b}(b)).

The simulated times $\tau_{T,SIM}$ show a maximum if the barrier
height equals the energy of the incident wave packet, i.e. $V_0
\approx E_0$. For higher barriers the traversal time becomes
smaller; smaller than the traversal time without barrier.

This fact can be interpreted in this way: that the mean
``velocity'' of the electron is greater in the case of a very high
and wide barrier than in the case of a free particle. But we must
remember, that up to now the formalism is non-relativistic and a
relativistic formalism would perhaps give different results.

The B\"uttiker Larmor approach and the ``semi-classical'' approach
show in the ranges $V_0 < 3 eV$ and $V_0 > 11 eV$ qualitatively
the same behavior as our simulation: the traversal times decrease
for increasing barrier height and are also smaller for very high
barriers than the time without barrier.

Moreover there is a dependence between the traversal and
reflection times and the detector parameter. The probability of an
ideal measurement in the standard theory of Quantum Mechanics does
not depend on the detector as these are only measurements of an
infinitesimal duration. For continues measurements, such a
dependence is not surprising.

\section{Quantum Chaos and Quantum Fractals} When we speak
about "chaos," we usually mean instability in the motion of most
classical systems; that is, system behavior that depends so
sensitively on the system's precise initial conditions that it is,
in effect, unpredictable and cannot be distinguished from a random
process. This kind of behavior is not to be expected in quantum
systems, essentially, for two different, yet related, reasons. The
first of these is that quantum evolution equations are linear; and
the second is that Heisenberg's indeterminacy smoothes out subtle
intricacies of classical chaotic orbits. The result is that there
are several different approaches to "quantum chaos". One approach
is to study the dynamics of quantum systems which are classically
chaotic; that is to study non-stationary states. Another approach
is to look at stationary states and concentrate on the form of the
wave function (or its Wigner distribution function). Yet another
approach is to concentrate on energies of stationary states, and
how the distribution of quantum energy eigenvalues reflects the
chaos of the classical trajectories (cf. \cite{mb}). Finally one
can discuss the problem of algorithmic inaccessability of certain
quantum mechanical states \cite{ap}.

The quantum chaos that we want to study has nothing to do with any
of the above. It is a new category, and it arose naturally out of
our approach to the quantum measurement problem. According to our
definition: Quantum Chaos is the chaotic behavior of quantum jumps
and accompanied readings of classical instruments in a particular
class of experiments, namely when experiments are set so as to
perform a simultaneous, continuous, fuzzy measurement of several
incompatible (i.e. incommeasurable, or noncommuting) observables.
This kind of behavior is easily modeled in EEQT, as EEQT is the
only theory (even if only semi-phenomenological) that provides
ways of simple mathematical modeling of "experiments" and
"measurements" on single quantum systems.

The example we present here, modeling measurement of spin
simultaneously in four different directions, was first introduced
in an unpublished report \cite{jad93e} by one of us (AJ), and then
given as a subject of PhD thesis to G. Jastrzebski
\cite{jas96}.\footnote{Another, extreme, example, which leads to a
random walk on a 2-sphere is discussed in Ref. \cite{jad94a}}

Before we describe the model, and the resulting chaotic behavior
and strange attractor on quantum state space, let us make first a
comment about the very question of simultaneous measurability of
noncommuting observables. This subject has become quite
controversial since the early formulation of Heisenberg's
uncertainty relations. Mathematically these relations are precise
and leave no doubt about their validity. But, the question of how
to interpret them physically and philosophically, has become a
subject of hot discussions. To quote from Popper's "Unended Quest"
\cite{popper}:
\begin{quotation} "The Heisenberg formula {\sl do not refer to measurements};
which implies that the whole current "quantum theory of
measurement" is packed with misinterpretations. Measurements which
according to the usual interpretation of the Heisenberg formulae
are "forbidden" are according to my results not only allowed, but
actually required for {\sl testing} these very formulae."
\end{quotation}

 Hilary Putnam came to a similar conclusion \cite{putnam81}:
\begin{quotation} "Recently I have observed that {\sl it
follows from just the quantum mechanical criterion for measurement
itself} that the "minority view" is right to at least the
following extent: simultaneous measurements of incompatible
observables {\sl can be made}\ . That such measurement cannot have
"predictive value" is true ..."
\end{quotation}

These words, written almost twenty years ago, suggest to us that
there is some chaotic behavior involved, and that this chaos and
its characteristics ought to be studied, both theoretically and
experimentally. Yet, for some reason, either no one noticed, or
they were not interested in looking into the problem
quantitatively. We need to ask why? Perhaps for the very same
reason that no one has been paying attention to the fact that {\bf
events do occur}. To quote from Tom Phipps' "Heretical Verities",
where he describes the publication of his paper "Do Quantum Events
Occur" in IEEE Journal:
\begin{quotation}"Recognizing that physics
and physicists were dead, I thought to determine if electrical
engineers were more alive. The answer was {\sl no}\ . I am
currently considering appealing the matter to an unbiased audience
of farmers ... "
\end{quotation}

Our present point of view on quantum events is rather similar to
that advocated by Phipps over twelve years ago. But, one needs
more than a point of view, and, fortunately, we also have a
precise mathematical model to deal with the subject in a
quantitative way.

 \subsection{Tetrahedral spin model} The model
was constructed to be as simple as possible, and yet interesting.
The simplest quantum system is spin $1/2$, which can be oriented
toward any point on a sphere. Mathematically we are dealing with
Hilbert space ${\bf C}^2$ of two complex dimensions. Quantum
states are rays in this space, thus elements of the projective
plane $P{\bf C}^2$, which is isomorphic to two-dimensional sphere
$S^2$. Equivalently, each quantum spin state can be thought of as
being an eigenstate of spin operator $\vec{\sigma }\cdot \vec{n}$
along some direction $\vec{n}$. To get a simple and yet
interesting behavior, we will couple our quantum spin to four
yes-no classical devices that are designed so as to make fuzzy
measurements of spin direction in four different space directions
simultaneously (and, it is no wonder that the resulting behavior
of our system will be, as we will soon see, somewhat
schizophrenic!).

Why did we take four spin directions rather than two or three?

Well, for simplicity we want our directions to be symmetrically
distributed. Two directions would point to south and north poles
of the sphere, and spin components along these directions commute
- thus no chaos.

Three symmetrically distributed directions would have to be
distributed along the equator, thus producing essentially
one-dimensional attractor.

 The simplest symmetric figure that uses all of three-dimensional
freedom, and thus produces an interesting two-dimensional
attractor, is a tetrahedron! And so we choose four unit vectors
${\vec{n}_i}$, $i=\,1,...,4$, arranged at the vertices of a
regular tetrahedron $$\vec{n}_1\;=\;(1,\,0,\,0),\;\;\vec{n}_2\;=
\;(-\frac{1}{3},\,0,\,\frac{2\sqrt{2}}{3})$$
$$\vec{n}_3\;=\;(-\frac{1}{3},\,\sqrt{\frac{2}{3}},\,-\frac{\sqrt{2}}{3}),\;\;
\vec{n}_4\;=\;(-\frac{1}{3},\,-\sqrt{\frac{2}{3}},\,-\frac{\sqrt{2}}{3})$$
Details of the dynamics of our model has been described elsewhere
\cite{blajaol99}. Here we will describe only the resulting
non-linear iterated function system, with point dependent
probabilities. The four nonlinear transformations acting on a
point $\vec{r}$ on the sphere are $$T_i:
\vec{r}\mapsto\vec{r}_i\;=\;\frac{(1\:-\:\alpha^2)\vec{r}\:
+\:2\alpha(1\:+\:\alpha\vec{r}\cdot\vni)\vni}{1\:+\: \alpha^2\:
+\:2\alpha\vec{r}\cdot\vni},\,\,\, i=1,\ldots ,4\;,$$ where
$0<\alpha < 1$ is a fuzziness parameter (in the limit $\alpha=1$
the measurements are sharp). At each step transformations $T_i$
are chosen with point dependent probabilities:
$$p_i(\vec{r})\;=\;\frac{1\:+\:
\alpha^2\:+\:2\alpha\vec{r}\cdot\vni}{4(1\:+\:\alpha^2)}.$$ Using
the above formulas it is easy to check that each $T_i$ indeed maps
unit sphere onto itself, that is that if $\vec{r}^2=1$ then also
$T_i(\vec{r})=1$, and also that $p_1+\ldots+p_4=1\ .$ Moreover,
each $T_i$ is one-one.

Computer simulations show that the resulting iterated function
system has a strange attractor whose fractal dimension decreases
from $1.44$ to $0.49$ when $alpha$ increases from $0.75$ to $0.95$
\cite{jas96}.

It should be noted that our iterated function system is not quite
of the usual type. Our maps $T_i$ are not contractions - $T_i$
contracts around the direction $\vni$, but acts as an expansion at
the opposite pole. Therefore the form of point-dependent
probabilities $p_i$ is important for convergence of the iteration
process.

\begin{figure}[p]
  \begin{center}
    \leavevmode
    \includegraphics[width=12cm, keepaspectratio=true]{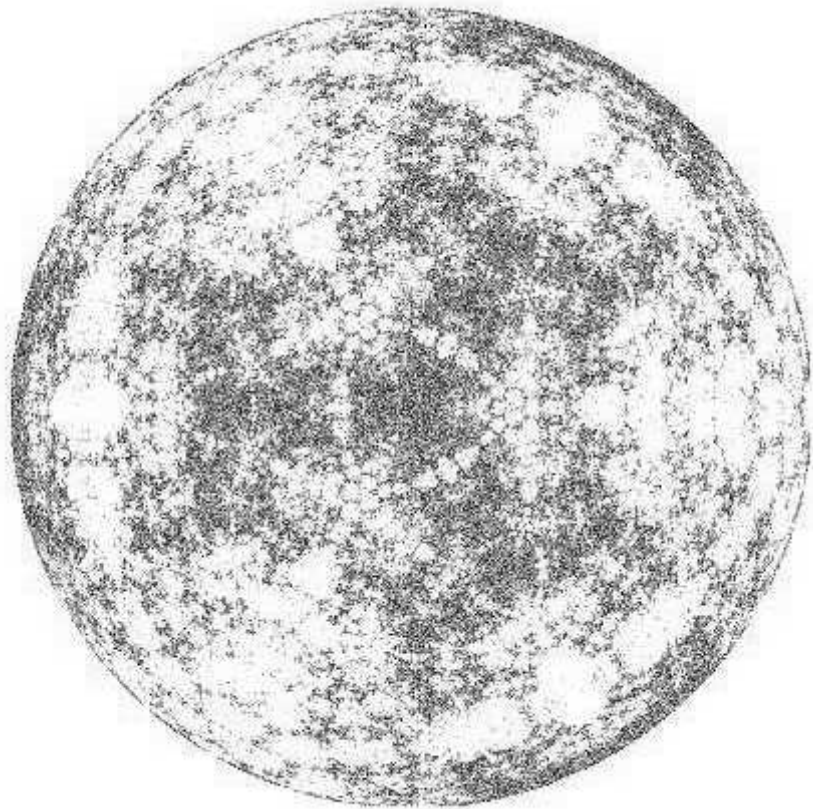}
  \end{center}
  \caption{Tetrahedral Quantum Fractal: Quantum state trajectory
  for $\alpha = 0.7$, $1000000000$ jumps.} \label{fig_4}
\end{figure}

There are two ways to visualize the attractor. The most evident
one, widely used for iterated function systems with contractive
affine maps, is to use the iteration process applied to some
initial vector $\vec{r}$. Fig.~\ref{fig_4} gives an illustration
of this method applied to our case. But, because of the fact that
the maps $T_i$ are, in our case, one-to-one and onto, we can apply
here another method, that is not applicable for affine iterated
function systems. This other method consists of iterations of the
associated Markov operator $P$ applied to some initial measure.
Invertibility of transformations $T_i$ assures then that if the
initial measure $\mu_0$ is Lebesgue continuous, then all
$\mu_n=P^n (\mu_0)$ are also Lebesgue continuous, and thus can be
easily visualized as functions $\mu_0(x)$ and $\mu_n(x)$
respectively.

\begin{figure}[p]
  \begin{center}
    \leavevmode
    \includegraphics[width=12cm, keepaspectratio=true]{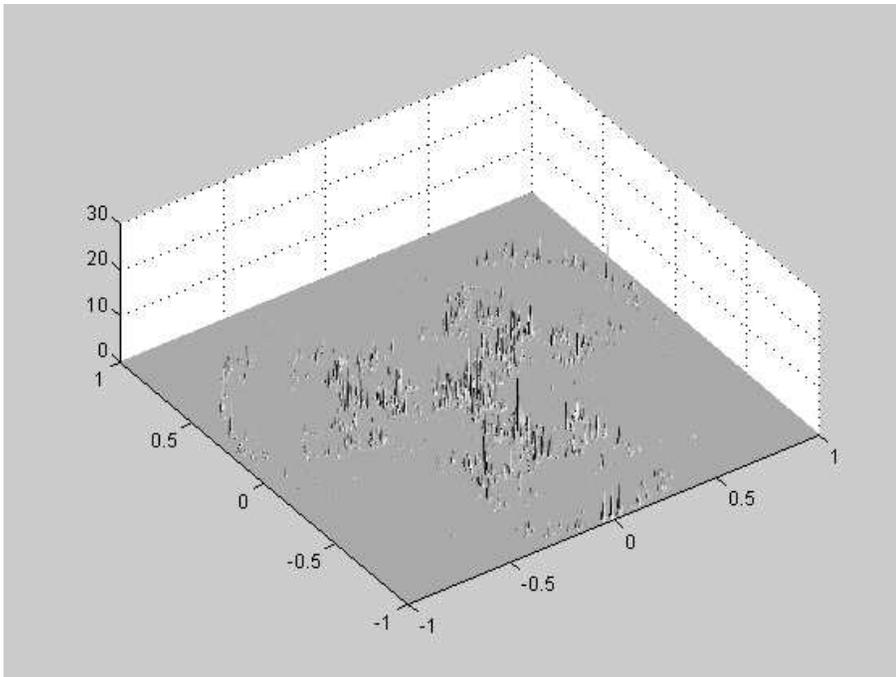}
  \end{center}
  \caption{Tetrahedral Quantum Fractal: $\alpha = 0.7$,
Approximation to the strange attractor. $8$th power of the Markov
operator applied to the uniform measure. Plane view of the upper
hemisphere. $Log(1+\mu_8(x))$ on the vertical scale. }
\label{fig_5}
\end{figure}

\begin{figure}[p]
  \begin{center}
    \leavevmode
    \includegraphics[width=12cm, keepaspectratio=true] {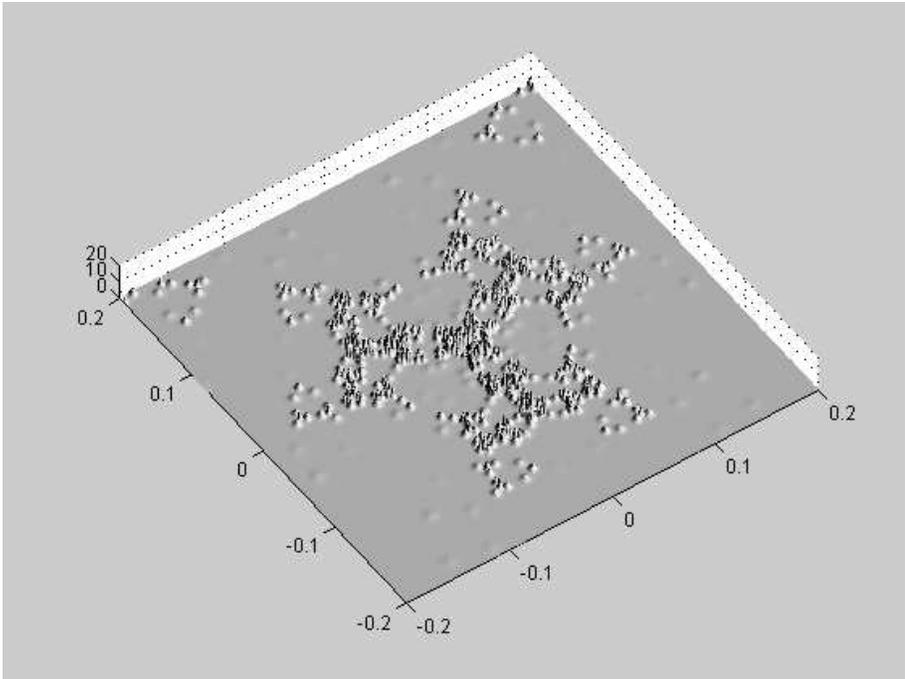}
  \end{center}
  \caption{Tetrahedral Quantum Fractal: $\alpha = 0.7$,
$5\times$zoom shows self-similarity. $Log(1+\mu_7(x)$ on the
vertical scale.} \label{fig_6}
\end{figure}

Fig.~\ref{fig_5} show 8th iteration of the Markov operator applied
to the invariant measure on the sphere (plane view of the upper
hemisphere), while Fig.~\ref{fig_6}, shows $5\times$ zoom of the
7th iteration .

\section{Concluding remarks}
In the foregoing examples, we have seen that EEQT is, indeed, an
enhancement of the standard quantum formalism, for the most
important reason that it allows us to discuss, in a quantitative
way, topics that are not easily treated within the orthodox
approach: time series of events generated by individual quantum
systems, generation of cloud chamber tracks, tunneling times,
simultaneous measurement of non-commuting observables, back-action
of classical variables on a quantum system, etc. EEQT can also
provide an engine powering Everett-Wheeler many-world
branching-tree.

In spite of all of these advantages and useful maneuvers, these
practical applications, EEQT is still not a fully developed
fundamental theory; though we are working in this direction. One
of the arbitrary factors we have to deal with is that the coupling
operators $\gab$ have to be cooked up in each case. In simple
cases, like those discussed in the present paper their choice is
rather unproblematic, yet even then we are not quite happy with
justification of this rather than another choice. One possible way
out would be to adhere to the often expressed point of view that
all measurements can be, in a final instant, reduced to position
measurements. Then, we can try to reduce every position
measurement to sharp Dirac delta-function detectors. Yet, even
then, we are left with an arbitrary value of a coupling constant
for each of the point detectors. This arbitrariness, although not
so much of a problem in practical applications of EEQT (for
instance, as shown in Ref. \cite{blaja96a} , for a wide range of
values of the coupling constant, change of its value affects only
the overall normalization constant), yet it makes us wonder about
the iceberg floating beneath the tip of EEQT that we DO see?

Frankly speaking we do not know. But, from all we do know,  we can
speculate about possible future evolution of EEQT. This
speculation goes back more than ten years, to a paper by one of us
\cite{jad90}, a paper which set up the program of which EEQT is a
partial realization. Quoting from this paper:
\begin{quotation}
The theory, the main idea of which we have just sketched, must
include into its scope two extremely different realities: the
classical world and the quantum world. Or, making the division in
a different direction: the world of matter, and the world of
information. However, the differences between these two aspects of
reality are so great, that their unification seems to be
impossible without a "catalyst", and we guess that this catalyst
is light. (...) Coherent infra-red photon states lead to
continuous superselection rules or, in other words, algebra of
observables of the photon field has a non-trivial center, whose
elements parameterize infra-red representations. (...) Classical
information is coded into the shape of infra-red photon cloud.
\end{quotation}
Thus one of our future projects is deriving EEQT from quantum
electrodynamics, where the classical parameter enters naturally as
the index of inequivalent non-Fock infrared representations. We
believe that using infinite tensor product representations of
quantum systems with an infinite number of degrees of freedom, we
will arrive naturally at our $g_\alpha\beta$ operators relating to
Hilbert spaces of inequivalent representations of CCR\/CAR.

\noindent {\bf Acknowledgments} One of us (A.J) would like to
thank L.K-Jadczyk for invaluable help.

%--------------- Fig. 1 --------------------
%------------- Fig. 2 ---------------------
%------------------- Fig. 3a+3b --------------------
%------------------- Fig. 4 --------------------
%------------------- Fig. 5 --------------------
%------------------- Fig. 6 --------------------

\end{document}